\documentclass[prc]{revtex4}

\usepackage{graphicx,amsmath}

\begin{document}

\title{Thermal production of the $\rho$ meson in the $\pi^+\pi^-$ channel }

\author{Scott Pratt}
\email{pratts@pa.msu.edu}
\affiliation{Department of Physics and Astronomy, Michigan State
University, East Lansing Michigan, 48824}
\author{Wolfgang Bauer}
\email{bauer@pa.msu.edu}
\affiliation{Department of Physics and Astronomy and National Superconducting 
Laboratory, Michigan State University, East Lansing Michigan, 48824}
\date{\today}

\begin{abstract}
\bigskip
Recent measurements of the $\pi^+\pi^-$ invariant mass distribution at RHIC
show a shifted peak for the $\rho$ meson in 100$A$ GeV in peripheral Au + Au 
and even in $p$ + $p$ collisions. A
recent theoretical study based on a picture of in-medium production
rates of pions, showed that a large shift could result from a combination of
the Boltzmann factor and the collisional broadening of the $\rho$.  Here
we argue that the two-pion density of states is the appropriate
quantity if one assumes a sudden break-up of the system. Methods for
calculating the density of states which include Bose effects are derived.  The
resulting invariant mass distributions are significantly enhanced at lower
masses and the $\rho$ peak is shifted downward by $\sim$ 35 MeV.

\end{abstract}

\maketitle

\section{Introduction}
\label{sec:intro} 

One of the most compelling motivations for studying heavy ion collisions is the
prospect for observing the restoration of chiral symmetry. The spontaneous
breaking of chiral symmetry is accompanied by the creation of a quark-antiquark
condensate whose coupling to nucleons is responsible for the great bulk of the
nucleon mass, and is therefore responsible for most of the mass of the
universe. The transient nature of the heavy-ion reaction precludes a detailed
investigation of all the quasi-particle modes in the highly excited collision
volume. However, the $\rho$ meson is unique for it typically decays inside the
spatial region where the vacuum structure might undergo novel changes. A
neutral $\rho$ decays with 99\% probability into a $\pi^+\pi^-$ pair and decays
with a small probability into $e^+e^-$ or $\mu^+\mu^-$ pairs. The
electromagnetic channels are especially useful because dilepton pairs will
largely leave the collision volume unscathed by interactions with the thousands
of other constituents. Since the $\rho$ has the same quantum numbers as the
photon, the invariant mass spectrum of dileptons is dominated by the $\rho$ for
masses between 600 and 800 MeV. Experiments at the CERN SPS for $e^+e^-$
\cite{ceres} and $\mu^+\mu^-$ \cite{na50,na38} suggest that the $\rho$ has
either dissolved \cite{siemenschin} (as would be expected in a quark-gluon
plasma), has moved down a few hundred MeV \cite{brownrho} (due to chiral
symmetry restoration), or has been broadened via collisions by many hundreds of
MeV \cite{rappwambaugh}.

Recently, the possibility of studying in-medium properties of the $\rho$ meson
through the $\pi^+\pi^-$ channel has been discussed
\cite{kolbprakash,rapp}. Unlike dileptons, pions are not penetrating
probes and are likely to re-interact before they escape. Since temperatures
fall to near 100 MeV at breakup, where the $\rho/\pi$ ratio falls to a few
percent, the chance that a $\pi^+$ is accompanied by a $\pi^-$ that originated
from the same $\rho$, rather than a charged pion from a different source, is
only a few percent. Thus, a background subtracted invariant mass distribution
should have a $\rho$ peak that comprises only a few percent of the integrated
distribution.

The STAR collaboration at RHIC has measured such a peak in $pp$
collisions, and for the first time, in peripheral relativistic heavy ion collisions
\cite{star_result}.  A surprisingly significant downward shift of the mass was observed
even in $pp$ collisions, especially at low $p_t$, and an even larger shift was
observed in peripheral Au + Au collisions. Results are not yet available for
central collisions where it is more difficult to observe the peak since the
$\rho/\pi$ ratio falls. Eventually, the $\rho$ peak should also be measured for
central collisions given sufficient statistics.

In references \cite{kolbprakash} and \cite{rapp}, the mass distribution was
predicted by considering the in-medium rate of $\rho$ decays into $\pi^+\pi^-$
pairs, $dN/dM d^3x dt$. This is the same approach as has been applied for
dilepton studies. In reference \cite{rapp}, these rates were corrected for
collision broadening and for Bose effects. Collision broadening was shown to be
particularly important in moving strength to lower-lying masses. However,
emission of pions is of a fundamentally different character than that of
dileptons. First, the final-state distribution is not necessarily proportional
to the decay rate since the decay rate is often balanced by a formation rate of
similar magnitude. Secondly, collisional broadening can not be applied in the
same manner since measurements are made in the asymptotic state. Finally, the
presence of the $\rho$ alters the two-pion scattering partial waves at non-zero
separations which should affect the mass distribution. As we will demonstrate,
the production-rate calculations of \cite{kolbprakash} and \cite{rapp} provide
different results than a freeze-out prescription which is governed by
the available phase space.

If the last strong interactions felt by the two pions used in the distribution
can be considered sufficiently hard to statistically sample the outgoing phase
space, the two-pion density of states should govern the invariant mass
distribution. Other non-randomizing interactions, identical particle
symmetrization and mean field interactions, would then serve to modify the
density of states. Although these $\rho$s probably decayed during the breakup
stage, which is well below the critical density, the decaying $\rho$ mesons
might still sample a region where mean-field effects, i.e., in-medium mass
shifts, are not negligible. Since pions are Goldstone bosons, they probably
leave the region with their energy and momenta unchanged during their exiting
trajectory, and one expects that a modification of the 2-pion invariant mass
distribution would reflect the in-medium modifications of the $\rho$ rather
than those of the pion. Whereas, this mass shift may be on the order of 100 MeV
at high temperature, it is unlikely to be much more than 25 MeV at breakup when
densities have fallen well below nuclear density.

It is not the aim of the current paper to model the in-medium mass shift of the
$\rho$, but rather to investigate how the invariant-mass distribution would
look in a thermal description based entirely on the two-pion density of states,
and the associated Bose effects. In the next section, methods for calculating
the two-particle density of states are presented along with a comparison with
the functional forms one would expect from rate calculations. After convoluting
with the Boltzmann weighting, we find that the $\rho$ peak is shifted downward
by $\sim 30$ MeV relative to the nominal $\rho$ mass. The shift is due to
three factors, the Boltzmann weighting \cite{barz}, the fact that the density
of states peaks below the $\rho$ mass, and the inclusion of other partial
waves. Bose-Einstein effects also enhance the distribution at lower masses
\cite{lafferty,opal,rapp}, especially for heavy ion collisions where the 
pionic phase space filling factors are approaching unity
\cite{bertsch,prattqm2002}.  In section \ref{sec:bose} methods are presented
for including Bose effects into the two-pion density of states. The resulting
mass distribution is strengthened at lower invariant masses, but the peak did
not shift appreciably.

\section{Invariant mass distributions from the two-pion density of states}
\label{sec:densityofstates}

Since the first measurements of the $\rho$ meson \cite{abolins,erwin}, the
masses and widths have fluctuated by several MeV depending on the analysis.
Currently, the Particle Data Group assigns a nominal mass of 771.1 MeV and a
width of 149.2 MeV
\cite{pdg}, with uncertainties for each number being near 1 MeV. The $\rho$
mass has been determined from a number of means, $e^+e^-\rightarrow \pi^+\pi^-$
reactions, $pp$ collisions, and $\pi p\rightarrow\pi\pi p$ reactions
\cite{protopopescu}. Electro-production of the $\rho$ is complicated by the
interference with the $\omega\rightarrow 2\pi$ channel
\cite{omegainterference} which constructively interferes with the $\rho^0$
channel since the electromagnetic coupling violates isospin conservation. Since
$pp$ collisions are typically highly inelastic, extracting the $\rho$ mass is
complicated by the same factors that complicate the study in a heavy-ion
environment. In the $\pi p\rightarrow \pi\pi p$ reaction, the proton is treated
as a source of pions which are assumed to scatter elastically with the incoming
pions. In fact, $\pi^+\pi^-$ phase shift analyses have been successfully
performed. The cross section for $\pi^+\pi^-$ reactions should have a
Breit-Wigner form,
\begin{equation}
\label{eq:sigma}
\sigma(M)=3\frac{4\pi}{q^2}\frac{(\Pi_I)^2}
	{(M^2-M_0^2)^2+(\Pi_I)^2},
\end{equation}
where $q$ is the momentum of either pion in the center-of-mass frame
($M=2\sqrt{m_\pi^2+q^2}$), and $\Pi_I$ is the imaginary part of the one-loop
self energy of the relativistic propagator.
\begin{eqnarray}
\label{eq:piself}
\Pi_I&=&\Pi_{I,0}\frac{M_0}{M}\left(\frac{q}{q_0}\right)^3F(q,q_0),\\
\Pi_{I,0}&=&\Gamma_0 M_0.\\
\end{eqnarray}
Here, $\Gamma_0$ is the nominal width, and $q_0$ is the momentum required to
provide the nominal mass, $M_0$. The last term, $F(q/q_0)$ is a form factor
whose exact form is in doubt\cite{lafferty}. The product $q^2\sigma$ peaks
precisely at the nominal mass irrespective of the form factor. 

The spectral function of the $\rho$ is related to the imaginary part of the
propagator,
\begin{eqnarray}
\label{eq:rhospectral}
S_\rho(M)&=&\frac{2M}{\pi}\Im \frac{1}{(M^2-M_0^2)+i\Pi_I},\\
&=&\left(\frac{2M}{\pi\Pi_I}\right)BW(M)\\
BW(M)&=&\frac{(\Pi_I)^2}{(M^2-M_0^2)^2+(\Pi_I)^2}.
\end{eqnarray}
Here, $S_\rho$ is usually associated with the number of states available to the
$\rho$ with a given mass. The real part of $\Pi$ is being ignored for the
current discussion. Since the Breit-Wigner function, $BW(M)$, always peaks at
$M=M_0$, and since $\Pi_I/M$ is rapidly growing with $M$ near the $\rho$ mass,
the spectral function always peaks below the $\rho$ mass. Setting the form
factor in Eq. (\ref{eq:piself}) to unity, the peak of the $\rho$ spectral
function shifts downward by 5 MeV. Applying some of the different expressions
for $\Pi_I$ discussed in \cite{lafferty} may result in the peak being shifted
further downward, perhaps as much as an additional 5 MeV.

The change in the total density of states can be expressed in terms of phase shifts,
\cite{pratt87,landaulifshitz}:
\begin{equation}
\Delta\rho(M)=\frac{1}{\pi}\sum_\ell(2\ell +1)\frac{d\delta_\ell}{dM}.
\end{equation}
Given the relation between the phase shift and the self energy, one can express
$\Delta\rho$ in terms of the self energy,
\begin{eqnarray}
\label{eq:rhoofdelta}
\tan\delta&=&\frac{\Pi_I}{M_0^2-M^2},\\
\Delta\rho_{\pi\pi}(M)&=&\frac{3}{\pi}\frac{2M \Pi_I}{(M_0^2-M^2)^2-\Pi_I^2}
\left(1+\frac{M_0^2-M^2}{2\Pi_I M}\frac{d\Pi_I}{dM}\right).
\end{eqnarray}
The first term is the spectral function of the $\rho$, which is often
associated with the probability of having a $\rho$ meson of mass $M$. Together,
the two terms describe the entire correction to the density of states,
including the effects of modifying the outgoing partial waves. 

\begin{figure}[t]
\centerline{\includegraphics[width=0.5\textwidth]{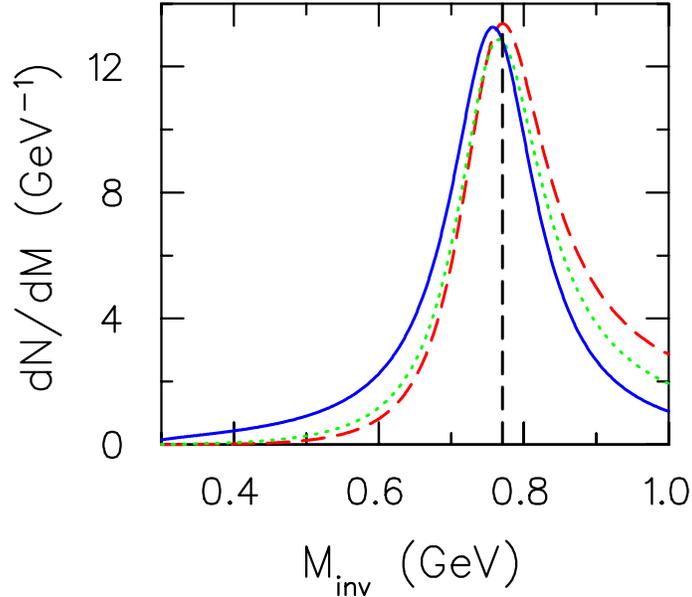}}
\caption{\label{fig:spectral}
The spectral function of the $\rho$ (dotted line) is broader than 
the Breit-Wigner form (dashed line). The total density of states, including
effects of modifying the outgoing partial waves, is noticeably shifted to the
left relative to the other forms. The difference is especially noticeable at
small invariant masses, where the three forms rise proportional to $q^6$, $q^3$
and $q$ respectively.}
\end{figure}

Figure \ref{fig:spectral} illustrates the importance of using the correct
expression for the density of states. The spectral function of the $\rho$ is
peaked below the Breit-Wigner function, and the total density of states is
peaked even lower. The difference is especially strong at low invariant masses,
as the Breit-Wigner function rises as $q^6$, the $\rho$ spectral function rises
as $q^3$ and the pionic density of states rises as $q$. This relative scaling
with $q$ would hold for any $p$-wave interaction.

\begin{figure}[t]
\centerline{\includegraphics[width=0.5\textwidth]{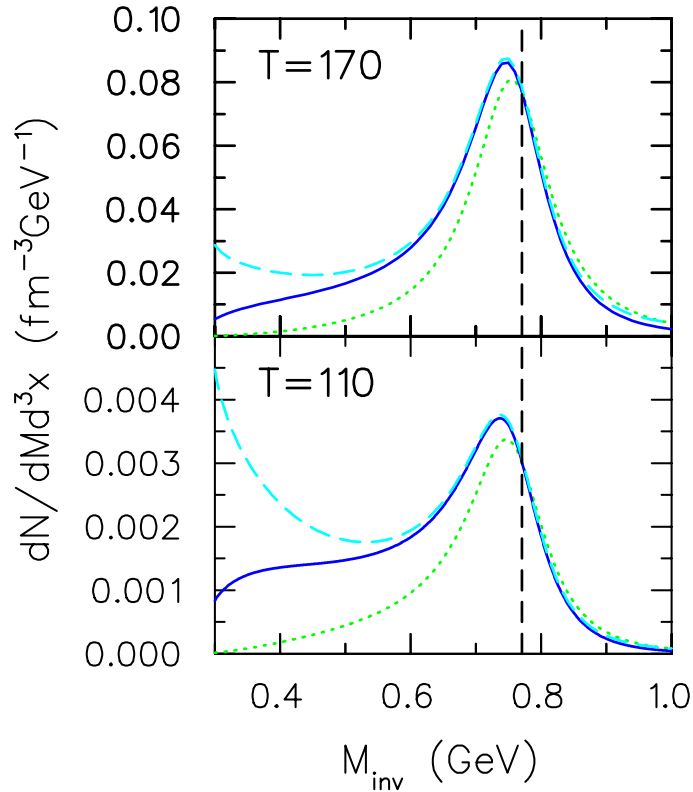}}
\caption{\label{fig:dndm}
The thermal mass distribution of $\pi^+\pi^-$ pairs is shown for three
calculations, both at $T=110$ MeV and $T=170$ MeV. Using the full density of
states as calculated from taking derivatives of phase shifts results in a
broader distribution for the $\ell=1$ channel (full line) than using the $\rho$
spectral function (dotted line). Including all $s$, $p$ and $d$ channels
(dashed line) provides significant strength at low invariant masses due to the
$s$ wave channels and moderate strength at higher masses from the $d$ wave
channels.}
\end{figure}

Thus far, the distribution of masses has not incorporated the Boltzmann factor,
which should push the peak even lower with the thermal weight, $e^{-M/T}$
\cite{barz}. More precisely, one needs to integrate over the modes in momentum
space due to relativistic effects,
\begin{equation}
\Delta \frac{dN_{\pi\pi}}{dMd^3x}=\int \frac{d^3P_\rho}{(2\pi)^3}
e^{-\sqrt{P_\rho^2+M^2}/T} \Delta\rho_{\pi\pi}(M).
\end{equation}
This should represent the background-subtracted 2-pion invariant-mass
distribution. As can be seen in Fig. \ref{fig:dndm}, the Boltzmann weight
pushes the distribution increasingly downward for lower temperatures. The upper
panel shows the mass distribution assuming a temperature of 170 MeV, which is a
reasonable temperature for thermal models of $pp$ collisions, while the lower
panel shows the result for a temperature of 110 MeV, which may be reasonable
for the breakup temperature in central heavy ion collisions. Calculations using
both the $\rho$ spectral function and the two-pion density of states are
displayed to illustrate the importance of choosing the appropriate form for the
density of states. The Boltzmann factor greatly magnifies the enhancements at
low $M$, to the point that a second peak appears for lower temperature.

The $\pi^+\pi^-$ density of states is also affected by phase shifts in other
channels. For $\pi^+\pi^-$, the $s$-wave channel is split into two isospin
components, 2/3 weight for $I=0$ and 1/3 weight for $I=2$. The $I=0$ channel is
particularly important as it corresponds to the mythical $\sigma$
meson. Although phase shift analyses do not reveal a sharp peak as in a
resonance \cite{pichowsky, kaminski, grayer,rosselet,shrinivasan}, the phase
shifts are considerable, rising steadily from zero at threshold to
approximately 90 degrees at $M=2M_K\sim 1$ GeV, where the kaon channel
opens. At the two-kaon threshold, the behavior of the phase shifts becomes
complicated and an inelastic treatment becomes warranted. Since one uses
derivatives of the phase shifts to find the density of states, interpolating
data for phase shifts can be dangerous due to noise in the experimentally
determined phase shifts. Thus, we apply a simple form that describes the
general behavior,
\begin{equation}
\delta_{I=0,S=0}=aq+b(M-2m_\pi).
\end{equation}
The first coefficient $a$ is the scattering length, which is small due to
constraints from chiral symmetry. The number varies throughout the literature
by several tens of percent. We use the value, $a=0.204/m_\pi$
\cite{kermani}. The second term does not contribute to the scattering
length, as $(M-2m_\pi)\sim q^2$ at low $q$. Choosing $b=9.1\times 10^{-4}$
GeV$^{-1}$ crudely reproduces experimental phase shifts, which are reviewed in
\cite{pichowsky}. Since these phase shifts rise half as far as those in the
delta channel, have one third the spin degeneracy, and have a 2/3 weight in the
$\pi^+\pi^-$ channel, they are noticeably less important than the $\rho$
channel in affecting the overall density of states, unless one is near the
two-pion threshold where $p$-wave interactions vanish.

Other phase shifts also contribute: $(I=2,\ell=0)$, $(I=0,\ell=2)$ and
$(I=2,\ell=2)$. Since none of these phase shifts exceed more than a few
degrees, they make nearly negligible contributions to the density of
states. For the $(I=2,\ell=0)$ channel, we apply an effective range expansion
\cite{losty},
\begin{equation}
\cot\delta=\frac{1}{qa}+\frac{1}{2}Rq,
\end{equation}
where $a=-0.13$ MeV$^{-1}$ and $R=1.0$ MeV$^{-1}$. The $d$ wave is also
composed of $I=0$ and $I=2$ pieces. For the $(I=0,\ell=2)$ piece, the data
\cite{estabrooks} are rough, and we make a simple expansion, 
\begin{equation}
 \delta_{I=0,\ell=2}=cq^5,
\end{equation}
where $c=6.2$ GeV$^{-1}$. The parameter $a$ is uncertain to the 50\% level. For
the $(I=2,\ell=2)$ partial wave, we use an expansion \cite{losty},
\begin{equation}
\delta_{I=2,\ell=2}=-8.4 q^5+12.5q^6 {\rm ~GeV}^{-1}.
\end{equation}
None of the these three channels are well understood, but none have a
substantial impact at or below the $\rho$ region of invariant mass.

Figure \ref{fig:dndm} also shows the invariant mass distribution of a thermal
ensemble with $T=110$ MeV using all the $s$, $p$ and $d$ channels. The $s$-wave
contributions are non-negligible near the $\rho$ mass, and dominate near the
two-pion threshold. The $d$-wave contributions matter only for masses near or
greater than 1.0 GeV.

\section{Bose Einstein Corrections}
\label{sec:bose}

Bose Einstein corrections should preferentially enhance low-mass pairs since
low-mass pairs are more likely to include a low-momentum pion. This has been
investigated within the context of the $\rho$ peak as well as the influence on
$Z$ boson decay modes \cite{lafferty}. In this section, we present a means to
include Bose enhancement effects which are consistent with the statistical
picture described in the previous section. 

In order to demonstrate Bose enhancement effects, we revert to the fundamental
definition of the two-particle density of states. 
\begin{eqnarray}
\label{eq:rhodef}
\rho(M)&=&\frac{1}{2\pi}\Im {\rm Tr}\frac{1}{M-H+i\epsilon}\\
&=&\frac{1}{2\pi}\Im {\rm Tr}\sum_{n=0}\frac{1}{M-H_0+i\epsilon}
\left(V\frac{1}{M-H_0+i\epsilon}\right)^n
\end{eqnarray}
We will work in the two-pion rest frame, so the trace would cover all two-pion
states that have total momentum zero. When including Bose effects, one would
sum all such two-pion states, plus average over the distribution of other
identical particles whose probability of being populated is
\begin{equation}
f({\bf q})=\frac{f_0({\bf q})}{1-f_0({\bf q})}=f_0({\bf q})(1+f({\bf q})).
\end{equation}
Thus, $(1+f)$ can be considered as an Bose enhancement factor while $f_0$ is
the phase space filling factor if Bose statistics were neglected.

Using the cyclic property of the trace, Eq. (\ref{eq:rhodef}) can be written in
terms of a derivative with respect to $M$,
\begin{eqnarray}
\rho(M)&=&\rho_0(M)+\frac{1}{2\pi}\Im \frac{d}{dM} {\rm Tr}
\sum_{n=1}\frac{1}{n}\left(\frac{V}{M-H_0+i\epsilon}\right)^n\\
&=&\rho_0(M)+\frac{1}{\pi}\Im \frac{d}{dM} {\rm Tr}
\sum_{n=1}\frac{1}{n}\left(\frac{{\cal P}}{M-H_0}V+i\pi\delta(M-H_0)V\right)^n.
\end{eqnarray}
One can expand the $n$ terms and note that the sum includes all possible
orderings of $n$ factors where each factor is either the principal value piece,
which is real, or the imaginary part which is proportional to the density of
states. One could restrict this sum to cover only those terms where the first
factor is the imaginary part and multiply by a factor of $n/N_\rho$, where
$N_\rho$ is the number of times that $i\pi\delta(M-H_0)$ appears in the
term. The sum over $n$ can then be transformed into a sum of all possible
numbers of appearances of the real part.
\begin{eqnarray}
\rho(M)&=&\rho_0(M)+\frac{1}{2\pi}\Im \frac{d}{dM} {\rm Tr}
\sum_{N_\rho=1}\frac{1}{N_\rho}
\left(i\pi\delta(M-H_0){\mathcal R}\right)^{N_\rho}\\
{\mathcal R}&\equiv&V+V\frac{{\cal P}}{E-H_0}{\mathcal R}.
\end{eqnarray}
Here, ${\mathcal R}$ is often referred to as the $R$-matrix. This can be written
in terms of a logarithm,
\begin{equation}
\Delta\rho(M)=\frac{1}{\pi}\Im\frac{d}{dM} {\rm Tr}
\log\left(\frac{1+i\pi\delta(M-H_0){\mathcal R}}{1-i\pi\delta(M-E){\mathcal R}}
\right).
\end{equation}
Thus, the density of states is determined completely by a single matrix,
\begin{equation}
\tau\equiv \pi\rho_0(M){\mathcal R},
\end{equation}
which is evaluated only for those states whose energy equals $M$.  In a
partial-wave basis, $\tau$ is related to the phase shift, $\tau=\tan\delta$. In
a plane-wave basis, the matrix $\tau$ links one direction of the relative
momentum with another, i.e., the matrix should be written with indices,
$\tau_{\Omega_1,\Omega_2}$.

The presence of other particles alters $\tau$. Each matrix element $V$ used to
construct $\tau$ is modified by the presence of other particles by the Bose
enhancement factor,
\begin{equation}
V({\bf q}_1,-{\bf q}_1;{\bf q}_2,-{\bf q}_2)\rightarrow
V({\bf q}_1,-{\bf q}_1;{\bf q}_2,-{\bf q}_2)
\sqrt{(1+f({\bf q}_1))(1+f(-{\bf q}_1))
(1+f({\bf q}_2))(1+f(-{\bf q}_2))}.
\end{equation}
If the intermediate states contained in the definition of ${\mathcal R}$ are
not affected by the phase space density, one can scale $\tau$ in the
same manner as $V$. Then, given the fact that each state appears in both the
bra and ket, one can modify $\tau$ in a simple manner to account for Bose
effects, 
\begin{equation}
\tau({\bf q}_1,-{\bf q}_1;
{\bf q}_2,-{\bf q}_2)=
\tau_0({\bf q}_1,-{\bf q}_1;
{\bf q}_2,-{\bf q}_2)
(1+f({\bf q}_1))(1+f(-{\bf q}_1)).
\end{equation}

The density of states is comprised of integrals of a cyclic nature, $I_n$,
\begin{eqnarray}
\Delta\rho(M)&=&\frac{1}{\pi}\Im\frac{d}{dM}\sum_{n=1,3,5\cdots} I_n/n\\
\label{eq:indef}
I_n(M)&=&\int \frac{d\Omega_1}{4\pi} \frac{d\Omega_2}{4\pi}\cdots 
\frac{d\Omega_n}{4\pi} \tau_0(\Omega_1,\Omega_2)\tau_0(\Omega_2,\Omega_3)
\cdots\tau_0(\Omega_n,\Omega_1)\\
\nonumber
&&(1+f({\bf q}_1))(1+f(-{\bf q}_1))\cdots (1+f({\bf q}_n))(1+f(-{\bf q}_n)).
\end{eqnarray}
Unless the momentum of the pair $P=0$, the phase space
densities will be sensitive to the direction of the relative momentum $\Omega$.

For a purely $s$-wave interaction, $\tau_0$ has no angular dependence and $I_n$
easily incorporates Bose effects,
\begin{equation}
I_n=\left(\tau_0 \int \frac{d\Omega}{4\pi} (1+f({\bf q}))
(1+f(-{\bf q}))\right)^n. 
\end{equation}
The correction to the density of states is then
\begin{eqnarray}
\Delta\rho(M)&=&\frac{1}{\pi}\frac{d\tau/dM}{1+\tau^2},\\
\tau&=&\tan\delta\int\frac{d\Omega}{4\pi} (1+f({\bf q}))
(1+f(-{\bf q})),
\end{eqnarray}
where $\delta$ is the phase shift as measured in the absence of Bose
modifications. 

For a $p$ wave interaction, $\tau_0$ has the angular dependence,
\begin{equation}
\tau_0(\Omega_1,\Omega_2)=\bar{\tau}_0 \hat{q}_1\cdot\hat{q}_2.
\end{equation}
By choosing a coordinate system where the $z$ axis is parallel to the total
pair momentum, there is reflection symmetry about the $x$, $y$ and $z$ planes. 
By making use of the identity,
\begin{eqnarray}
\int \frac{d\Omega_b}{4\pi} (\vec{A}\cdot\hat{b})
	(\hat{b}\cdot\vec{C}) F(\Omega_b)&=&\vec{A}'\cdot\vec{C}\\
A_i^\prime&=&A_iF_i,\\
F_x&=& \int \frac{d\Omega}{4\pi} F(\Omega) \cos^2\phi\sin^2\theta,\\
F_y&=& \int \frac{d\Omega}{4\pi} F(\Omega) \sin^2\phi\sin^2\theta,\\
F_z&=& \int \frac{d\Omega}{4\pi} F(\Omega) \cos^2\theta,
\end{eqnarray}
one can iteratively perform the integral in Eq. (\ref{eq:indef}).
\begin{eqnarray}
I_n(M)&=&\left(\bar{\tau}_0F_x\right)^n
+\left(\bar{\tau}_0F_z\right)^n+\left(\bar{\tau}_0F_z\right)^n,\\
\end{eqnarray}

Using $F(\Omega)=(1+f({\bf q}))(1+f(-{\bf q}))$, one can calculate
$\Delta\rho$,
\begin{eqnarray}
\Delta\rho(M)&=&\frac{1}{\pi}\left(\frac{d\tau_x/dM}{1+\tau_x^2}
+\frac{d\tau_y/dM}{1+\tau_y^2}+\frac{d\tau_z/dM}{1+\tau_z^2}\right),\\
\tau_x&=&\tan\delta\int \frac{d\Omega}{4\pi} (1+f({\bf q}))
(1+f(-{\bf q})) 3\sin^2\phi\sin^2\theta,\\
\tau_y&=&\tan\delta\int \frac{d\Omega}{4\pi} (1+f({\bf q}))
(1+f(-{\bf q})) 3\cos^2\phi\sin^2\theta,\\
\tau_z&=&\tan\delta\int \frac{d\Omega}{4\pi} (1+f({\bf q}))
(1+f(-{\bf q})) 3\cos^2\theta.
\end{eqnarray}
The calculation of $\Delta\rho(M)$ must be repeated for each value of the total
momentum since $f({\bf q})$, which is defined in the two-pion rest frame,
changes when the total momentum is changed.

The $p$-wave and $s$-wave corrections to the density of states do not interfere
with one another since they have opposite parities and $(1+f({\bf
q}))(1+f(-{\bf q}))$ has even parity. However, calculation of the $\ell=2$
contributions would be complicated by the fact that the elliptical distortion
of the Bose enhancement factors would mix the $\ell=0$ and $\ell=2$
contributions. For the calculations here, the $\ell=2$ contributions were
calculated by assuming that the Bose enhancement factors were independent of
$\Omega$, then using enhancement factors which had been averaged over all
directions of $\Omega$.

\begin{figure}[t]
\centerline{\includegraphics[width=0.5\textwidth]{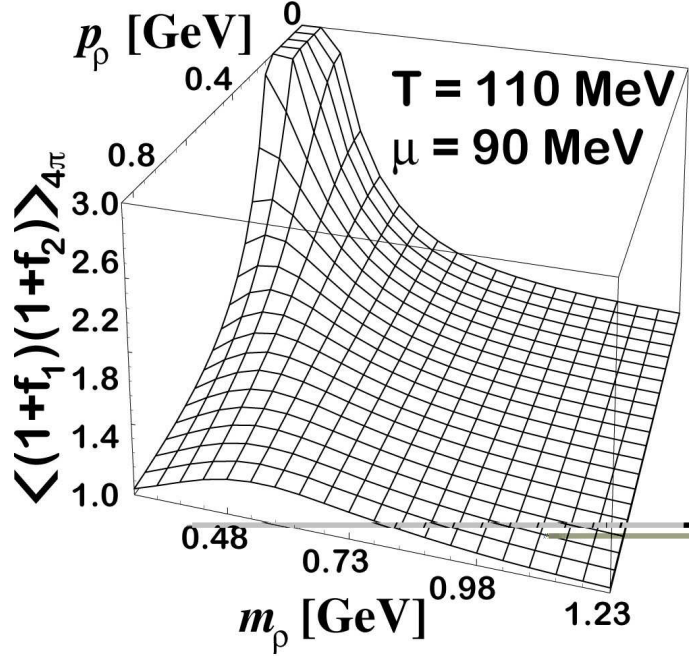}}
\caption{\label{fig:meanenhancement}
The mean values of $\langle (1+f_1)(1+f_2)\rangle$ are shown as a function of
the invariant mass and total momentum of the outgoing pion pair. The values
have been averaged over all directions of the relative momentum. The
enhancement factors exceed 2.0 for a low values of the pair momentum and
invariant mass.}
\end{figure}

The mean Bose enhancement, $\langle (1+f_1)(1+f_2)\rangle$, is shown as a
function of the invariant mass and momentum of the decaying $\rho^0$ in
Fig. \ref{fig:meanenhancement} assuming a breakup temperature of 110 MeV and an
effective chemical potential of 90 MeV. The enhancement has been averaged over
the directions of the relative momentum. The enhancement is largest for
low-momentum, low-mass pairs since these pions most strongly sample the region
of high phase space density. For higher invariant masses, the Bose enhancement
is actually stronger for higher pair momenta, as it allows one of the outgoing
pions to have low $p_t$ and sample the high phase-space density region. From
viewing Fig. \ref{fig:meanenhancement}, it is clear that the Bose modifications
to the invariant mass distribution would be more acute if experiments were to
focus on pion pairs with low total momentum.

\begin{figure}[t]
\centerline{\includegraphics[width=0.5\textwidth]{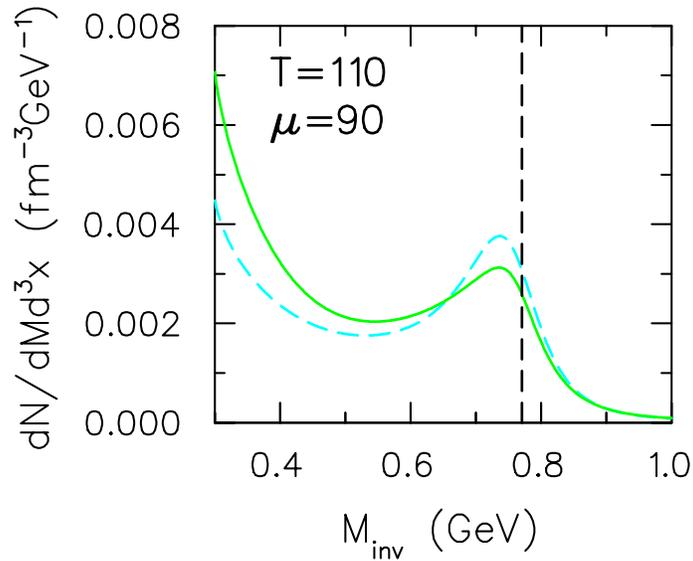}}
\caption{\label{fig:dndm_bose}
The mass distribution is shown with (solid line) and without (dashed line) Bose
effects for $T=110$ MeV, $\mu=90$ MeV. Bose effects enhance the probability of
producing low invariant-mass pairs since they are more likely to have low
momenta and stronger Bose enhancement factors.}
\end{figure}

Bose corrected densities of states are shown in Fig. \ref{fig:dndm_bose} for
$T=110$ MeV and $\mu=90$ MeV. A non-zero chemical potential was used to account
for the relative overpopulation of pionic phase space which may result from
rapid cooling \cite{gong} and might be magnified by the effects of chiral
symmetry restoration \cite{haglin}. Analyses of $\pi\pi$ correlations from RHIC
indeed point to high phase space densities
\cite{bertsch,prattqm2002}, especially for central collisions of heavy
ions. As expected, lower-mass states were more
enhanced by Bose effects. Since the density of states was proportional to the
derivative of $(\tan\delta\langle(1+f)(1+f')\rangle$, and since the averaged
phase space filling factors generally fall as $M$ increases, the density of
states was less enhanced for intermediate masses as compared to the no-Bose
case. The peak of the distribution shifted downward by only one MeV after the
inclusion of Bose effects.

Although the position of the peak was not much affected by Bose effects shown
in Fig. \ref{fig:dndm_bose}, Bose effects led to a near doubling of the
distribution at low masses. These effects are most important at low $p_t$ where
the phase space densities are higher. In $pp$ collisions, a movement of the
$\rho$ peak was observed for low $p_t$ pairs\cite{aguilarbenitez,star_result}
which is suggestive of Bose effects. However, at the $\rho$ peak each pion has
a relative momentum of $\sim 300$ MeV/c and will largely sample phase-space
regions with moderate to low phase space densities. Although the invariant mass
distribution is mainly altered at invariant masses below the $\rho$ peak, Bose
effects should contribute to washing out the peak by increasing the declining
background to the peak structure in Fig. \ref{fig:dndm_bose}.

\section{summary}

Our principal finding is that the $\rho$ peak in the $\pi^+\pi^-$ invariant
mass distribution should be approximately 35 MeV lower than the nominal $\rho$
mass, if one accepts the scenario of a sudden breakup that thermally samples
the two-pion density of states. The shift was the result of convoluting the
density of states which is shifted by $\sim$ 10 MeV below the $\rho$ mass with
the Boltzmann factor. Given the extra cooling inherent to heavy ion collisions,
the breakup temperature is probably near 110 MeV, well below the characteristic
temperatures used to describe $pp$ collisions. This low temperature is
responsible for the additional downward shift of the peak in heavy ion
collisions. In addition to the shift of the peak, the distribution showed
significant additional strength at invariant masses near the two-pion
threshold. This additional strength hinged on using the correct expressions for
the density of states, especially in the $\ell=0$ channels. Although the
position of the peak was not much affected by Bose effects, Bose effects led to
a near doubling of the distribution at low masses.

The thermal model presented here rests critically on a pair of
assumptions. First, we have assumed that the breakup is sudden, i.e., the last
strong interaction experienced by the particles samples the outgoing
two-particle phase space. Indeed, interferometric measurements do suggest a
sudden breakup \cite{starhbt,phenixhbt,prattqm2002}. If emission were gradual,
e.g., surface evaporation, this picture would be invalid. An appropriate
treatment of the surface would include the dynamics of surface penetration and
absorption and might include collisional broadening. For instance, spectral
lines in stars are affected by collisional broadening. The ``truth'' of the
breakup at RHIC probably has elements of both volume-like breakup and
surface-like evaporation. Thus, the effect of collisions, which played a
pivotal role in moving the distribution downward in
\cite{rapp}, requires more study.

The second assumption inherent to these calculations is related to the neglect
of finite-size effects. The enhancement factors applied to small-angle
correlation studies are usually based on the outgoing wave function,
$|\phi({\bf q},{\bf r})|^2$ \cite{bauergelbkepratt,heinzjacak}. For large
sources, there is a straight-forward correspondence between the integrated wave
functions and the phase shifts
\cite{boal},
\cite{corrtail},
\begin{equation}
\frac{1}{\pi}\sum_\ell (2\ell+1)\frac{d\delta_\ell}{dE}
=\frac{qM}{8\pi^2}\int d^3r
\left( |\phi({\bf q},{\bf r})|^2-|\phi_0({\bf q},{\bf r})|^2 \right).
\end{equation}
Since the modification of the wave function is mostly confined to a region
where $qR<\pi$, one expects that the treatments shown here should work well for
$q>100$ MeV/c, or for masses greater than 400 MeV. For masses near threshold, a
different approach, based on the actual scattered wave functions, would be
warranted. Such an approach could also incorporate the effects of the Coulomb
interaction between pions.

Finally, it should be emphasized that other correlations, besides those
resulting from the change in the two-pion density of states, will play a role
in any experimental measurement. Experimental analyses are typically based on a
like-sign subtraction. This should eliminate global correlations such as
elliptic flow which correlate same-sign and opposite-sign pairs
equally. However, any correlation based on charge conservation should survive
the subtraction
\cite{bassdanpratt,starbalance}. For every $\pi^+$, there is a $\sim 75$\% 
chance that local charge conservation will result in an extra $\pi^-$ being
emitted with a similar rapidity. This should provide a bump in the
like-sign-subtracted invariant-mass distribution that peaks for masses near 400
MeV. The ratio of the $\rho$ peak in the like-sign subtracted distribution to
the bump from charge correlation is approximately determined by the chance that
a given $\pi^+$ had its last interaction with other hadrons through the decay
of a $\rho^0$. This ratio should be smaller for central collisions since the
breakup temperature is lower which reduces the $\rho/\pi$ ratio.

The in-medium mass of the $\rho$ might be altered by $\sim$20 MeV at breakup.
Given that this peak is also spread out and distorted as shown in the
calculations presented here, it is certainly challenging to isolate the
contribution from the $\rho$ and to quote a peak height to a better accuracy
than 20 MeV. Upcoming runs at RHIC may increase the statistics by more than an
order of magnitude. Thus, we believe that there remains a good chance that the
$\rho$ can be studied in detail, even in the central collision Au+Au environment.

Finally, we compare the experimentally observed mass shifts to results of our
model. In reference \cite{star_result} it was reported that the $\rho$ shift
downward in $pp$ collisions by $\sim 20$ MeV at higher $p_t$ while shifting
downward by $\sim 45 MeV$ at low $p_t$. The shift appeared to be 5 to 10 MeV
larger for high-muliplicity $pp$ collisions and perhaps another 5 MeV lower for
peripheral Au+Au reactions. Similar behavior for $pp$ collisions had been
reported for $\sqrt{s}=27$ GeV $pp$ collisions
\cite{aguilarbenitez}. The shifts that we extracted were as large 35 MeV, but
these calculations assumed a lower temperature, 110 MeV, and a higher effective
chemical potential, 90 MeV, than would be appropriate for $pp$ phenomenology.
For a temperature of 170 MeV, and zero chemical potential, the shift was in the
range of 20 MeV, a somewhat smaller shift than what was observed by STAR. It
appears that the experimental mass shift is 10 to 20 MeV stronger than what we
would expect from our approach. But, before this discrepancy can be attributed
to novel in-media phenomena, i.e., a mass shift of the $\rho$, it should be
stressed that systematic uncertainties described in \cite{star_result} are of
the order of 10 MeV. This problem would be served well by both a higher
statistics experimental analysis and a more detailed theoretical modeling. An
improved calculation would consider finite-size effects, the influence of other
resonances, and the effects of experimental acceptances and efficiencies.

\begin{acknowledgments}
This work was supported by the U.S. National Science Foundation, Grant No.
PHY-02-45009 and by the U.S. Department of Energy, Grant No. DE-FG02-03ER41259.
\end{acknowledgments}


\begin{thebibliography}{99}
\bibitem{ceres} D. Adamova et al., http://arXiv.org, nucl-ex/0209024 (2002).
\bibitem{na50} NA50 Collab., M.C. Abreu et al., European Phys. J. C{\bf 13}, 
69 (2000). 
\bibitem{na38} NA38 Collab., M.C. Abreu et al., Phys. Lett B{\bf 368}, 239
(1996). 
\bibitem{siemenschin} P.J. Siemens and S.A. Chin, Phys. Rev. Lett. {\bf 55},
1266 (1985).
\bibitem{brownrho} G.E. Brown and M. Rho, Phys. Rep. {\bf 363}, 85 (2002).
\bibitem{rappwambaugh} R. Rapp and J. Wambach, Adv. Nucl. Phys. {\bf 25}
(2000). 
\bibitem{kolbprakash} P.F. Kolb and M. Prakash, Phys. Rev. C{\bf 67}, 044902
(2003).
\bibitem{rapp} R. Rapp, http://arXiv.org, hep-ph/0305011 (2003).
\bibitem{star_result} J. Adams et al., www.arXiv.org, nucl-ex/0307023 (2003).
\bibitem{barz} H.W. Barz, G. Bertsch, B.L. Friman, H. Schultz and S. Boggs, 
Phys. Lett. B{\bf 265}, 219 (1991).
\bibitem{lafferty} G.D. Lafferty, Z. Phys. C{\bf 60}, 659 (1993).
\bibitem{opal} P.D. Acton et al., Phys. Lett. B{\bf 267}, 143 (1991).
\bibitem{bertsch} G.F. Bertsch, Phys. Rev. Lett. {\bf 72}, 2349 (1994); 
{\em ibid.} {\bf 77}, 789(E) (1996).
\bibitem{prattqm2002} S. Pratt, Nucl. Phys. A{\bf 715}, 389c (2003).
\bibitem{abolins} M. Abolins, R.L. Lander, W. Mehlhop, N. Xuong 
and P.M. Yager , Phys. Rev. Lett. {\bf 11}, 381 (1963).
\bibitem{erwin} A.R. Erwin, R. March, W.D. Walker and E. West, Phys. Rev. Lett.
{\bf 6}, 628 (1961).
\bibitem{pdg} Particle Data Group, K. Hagiwara et al., Phys. Rev. D{\bf 66},
010001 (2002).
\bibitem{protopopescu} S.D. Protoposecu, M. Alston-Garnjost,
A. Barbaro-Galtieri, S.M. Flatt, J.H. Friedman, T.A. Lasinski, G.R. Lynch,
M.S. Rabin, and F.T. Solmitz, Phys. Rev. D.{\bf 7}, 1279 (1973).
\bibitem{omegainterference} P.S. Biggs et al., Phys. Rev. Lett. {\bf 24}, 
1201 (1970).
\bibitem{pratt87} S. Pratt, P.J. Siemens and Q.N. Usmani, 
Phys. Lett. {\bf B189}, 1 (1987).
\bibitem{landaulifshitz} L.D. Landau and E.M. Lifshitz, 
\underline{Statistical Physics}, Reed Educational and Professional Publishing,
Oxford, 230 (1999).
\bibitem{pichowsky} M.A. Pichowsky, A. Szczepaniak and J.T. Londergan, 
Phys.Rev. D{\bf 64}, 036009 (2001).
\bibitem{kaminski} R. Kaminiski, L. Lesniak and K. Rybicki,
Acta. Phys. Polon. B{\bf 31}, 895 (2000).
\bibitem{grayer} G. Grayer, et al., Nucl. Phys. B{\bf 75}, 189 (1974).
\bibitem{rosselet} L. Rosselet, et al., Phys. Rev. D{\bf 15}, 574 (1977).
\bibitem{shrinivasan} V. Shrinivasan et al., Phys. Rev. D{\bf 12}, 681 (1975).
\bibitem{kermani} M. Kermani et al., Phys. Rev. C{\bf 58}, 3431 (1998).
\bibitem{losty} M.J. Losty, V. Chaloupka, A. Ferrando, L. Montanet, E. Paul,
D. Yaffe, A. Zieminski, J. Alitti, B. Gandois and J. Louie, 
Nucl. Phys. B{\bf 69}, 185 (1974).
\bibitem{estabrooks} P. Estabrooks and A.D. Martin, Nucl. Phys. B{\bf 79}, 
301 (1974).
\bibitem{gong} C. Greiner, C. Gong and B. M\"{u}ller, Phys. Lett. B{\bf 16},
226 (1993).
\bibitem{haglin} S. Pratt and K. Haglin, Phys. Rev. C{\bf 59}, 3304 (1999).
\bibitem{aguilarbenitez} M. Aguilar-Benitez et al., Z. Phys. C {\bf 50}, 
405 (1991).  
\bibitem{starhbt}  C. Adler et al. Phys. Rev. Lett. 87 , 082301 (2001).
\bibitem{phenixhbt} K. Adcox, et al., Phys. Rev. Lett. {\bf 88}, 192302 (2002).
\bibitem{bauergelbkepratt} W. Bauer, C.K. Gelbke and S. Pratt, Ann. Rev. of
Nucl. and Part. Science {\bf 42}, 77 (1992).
\bibitem{heinzjacak} U. Heinz and B. Jacak, Ann. Rev. Nucl. Sci. {\bf 49}, 529 (1999).
\bibitem{boal} B. Jennings, D. Boal and J. Shillcock, Phys. Rev. C{\bf 33},
1303 (1986).
\bibitem{corrtail} S. Pratt and S. Petriconi, http://arXiv.org, 
nucl-th/0305018 (2003).
\bibitem{bassdanpratt} S.A. Bass, P. Danielewicz, and S. Pratt,
Phys. Rev. Lett. {\bf 85},  2689 (2000).
\bibitem{starbalance} J. Adams et al., Phys. Rev. Lett. {\bf 90}, 172301
(2003).  
\end{thebibliography}
\end{document}